\documentclass{moriond}
\newcommand\f[2]{\frac{#1}{#2}} 
\DeclareRobustCommand{\alphas}{\ensuremath{\alpha_{\mathrm{s}}} } 
\DeclareRobustCommand{\as}{\alphas} 
\DeclareRobustCommand{\mur}{\ensuremath{\mu_{\mathrm{R}}} }
\DeclareRobustCommand{\muR}{\mur}
\DeclareRobustCommand{\muf}{\ensuremath{\mu_{\mathrm{F}}} }
\DeclareRobustCommand{\muF}{\muf}
\DeclareRobustCommand{\qt}{\ensuremath{q_T} } 
\DeclareRobustCommand{\qtcut}{\ensuremath{q_T^\mathrm{cut}} } 

\def\be{\begin{equation}}
\def\ee{\end{equation}}
\def\bea{\begin{eqnarray}}
\def\eea{\end{eqnarray}}

\newcommand{\Photo}{\includegraphics[height=35mm]{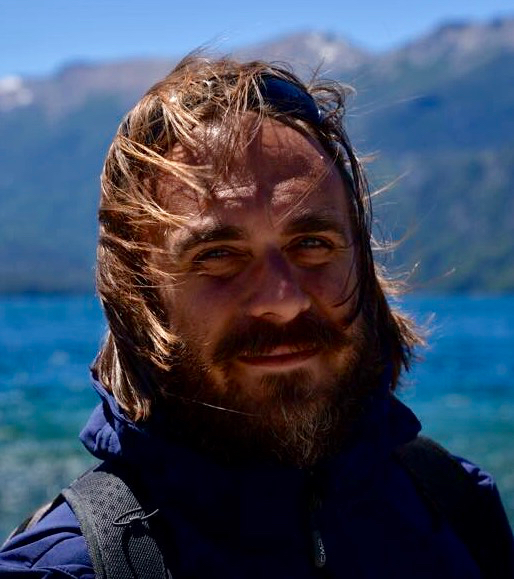}}
\begin{document}
\vspace*{4cm}
\title{THE TRANSVERSE MOMENTUM SUBTRACTION METHOD AT N$^3$LO APPLIED TO HIGGS BOSON PRODUCTION AT THE LHC}

\author{ L. CIERI }

\address{INFN, Sezione di Milano-Bicocca,
Piazza della Scienza 3, I-20126 Milano, Italy}

\maketitle\abstracts{
We consider the extension of the transverse--momentum ($\qt$) subtraction method at next-to-next-to-next-to-leading order (N$^3$LO) in perturbative QCD. While all the $\qt$-subtraction ingredients at $\qt \neq 0$ are known in analytical form, the third-order collinear functions and helicity-flip functions, which contribute only at $\qt=0$, are approximated using a prescription which uses the known result for the total Higgs boson cross section at this order. As a first application of the third-order $\qt$-subtraction method, we present the N$^{\rm3}$LO rapidity distribution of the Higgs boson at the LHC.
}
\section{Introduction}
\label{sec:intro}
Measurements at the LHC present an impressive and continuously improving quality, making even the next-to-next-to-leading order (NNLO) QCD perturbative accuracy not sufficient to match the demands of the LHC data. Processes which manifest such necessity to be computed beyond NNLO, exhibit next-to-leading order (NLO) corrections comparable in size with the leading order (LO), and where their NNLO corrections still exhibit large effects such that the size of the theoretical uncertainties remains larger than the experimental uncertainties. 

This motivated a new theoretical effort to go beyond NNLO, in order to include the next perturbative term: the next-to-next-to-next-to-leading order (N$^{\rm3}$LO). Sum rules, branching fractions~\cite{Chetyrkin:1994js} and deep inelastic structure functions~\cite{Vermaseren:2005qc} have been known to this order for quite some time. At present, the only hadron collider observables for which N$^{\rm3}$LO QCD corrections have been calculated are the total cross section for Higgs boson production in gluon fusion~\cite{Anastasiou:2015ema,Mistlberger:2018etf}, $b\bar{b}$ fusion \cite{Duhr:2019kwi}, in vector boson fusion~\cite{Dreyer:2016oyx} and Higgs boson pair production~\cite{Dreyer:2018qbw} in vector boson fusion. Recently, first steps have been taken towards more differential observables by computing several N$^{\rm3}$LO threshold expansion terms to the Higgs boson rapidity distribution in gluon fusion~\cite{Dulat:2017prg,Dulat:2018bfe}. In addition, the projection-to-Born method has been recently extended to N$^{\rm3}$LO~\cite{Currie:2018fgr} for jet production in deep inelastic scattering. 

In this proceeding we present the first extension of the $\qt$-subtraction method \cite{Catani:2007vq} at N$^{\rm3}$LO \cite{Cieri:2018oms} and we will apply it, to compute Higgs boson production differentially in the Higgs boson rapidity at N$^{\rm3}$LO accuracy. The proceeding is organized as follows: in Sec.~\ref{sec:forma} we recall briefly the main ideas of the $\qt$-subtraction formalism and we present the necessary ingredients up to  N$^{\rm3}$LO, specifying which elements are known analytically and identifying the missing coefficients at N$^{\rm3}$LO. Results for the N$^{\rm3}$LO Higgs boson rapidity distribution and the associated theoretical uncertainty is presented in Sec.~\ref{sec:results}. Finally, in Sec.~\ref{Sec:conclu} we summarize our results.
\section{The $\qt$-subtraction formalism at N$^{\rm3}$LO}
\label{sec:forma}
We consider the inclusive hard scattering reaction 
\begin{equation}
  h_1(p_1)+h_2(p_2)\to F(\{q_i\})+X\, ,
  \label{eq:class}
\end{equation}
where $h_1$ and $h_2$ denote the two hadrons which collide with momenta $p_1$ and $p_2$ producing the identified colourless final-state system $F$, accompanied by an arbitrary and undetected final state $X$. The colliding hadrons have centre-of-mass energy $\sqrt s$, and are treated as massless particles $s= (p_1+p_2)^2 = 2p_1\cdot p_2 \;$. The observed final state $F$ consists of a generic system of non-QCD partons composed of \emph{one or more} colour singlet particles (such as vector bosons, photons, Higgs bosons, Drell--Yan (DY) lepton pairs and so forth) with total invariant mass $M$ ($M^2=q^2 \;$), transverse momentum $\qt$ with respect to the direction of the colliding hadrons, and rapidity in the centre-of-mass system of the hadronic collision, $Y$ ($Y = \frac{1}{2} \ln \left(\frac{p_2\cdot q}{p_1\cdot q}\right) \;$). 

Our strategy is based on the following steps:  we first note that, at LO, the transverse momentum 
${\bf q}_{\, T}= \sum_i {\bf q}_{\, Ti}$ of the triggered final state $F$ is exactly zero. As a consequence, as long as $q_T\neq 0$, the N$^{i}$LO contributions \footnote{N$^{i}$LO stands for LO if $ i = 0$, NLO if $ i = 1$, NNLO for $ i = 2$ and finally, N$^{\rm3}$LO for $ i = 3$ obviously.} are actually given by the N$^{i-1}$LO contributions to the triggered final state $F+{\rm jet(s)}$.

Therefore, the cross section can be written as $d{\sigma}^{F}_{ \rm N^{i}LO}|_{q_T\neq 0}=d{\sigma}^{F+{\rm jets}}_{\rm N^{i-1}LO}$, implying that, in the limit $q_T\neq 0$, the infrared (IR) divergences in our N$^{i}$LO calculation are those in $d{\sigma}^{F+{\rm jets}}_{\rm N^{i-1}LO}$. Since we are interested in N$^{\rm3}$LO cross sections, NNLO IR singularities can be handled and cancelled by using available NNLO formulations of  subtraction methods (in our case, antenna subtraction \cite{Antenna:method}).

The only remaining singularities of N$^{3}$LO type are associated to the limit $q_T\to 0$, and we treat them by an additional subtraction ($\qt$-subtraction method at N$^{3}$LO) presented in Ref. \cite{Cieri:2018oms}.

The following sketchy presentation is illustrative; for more details, we recommend the foundational papers \cite{Catani:2007vq,Bozzi:2005wk,Cieri:2018oms}. We use a shorthand notation that mimics the notation of Ref.~\cite{Bozzi:2005wk}. We define our counter term \footnote{The symbol $\otimes$ understands convolutions over momentum fractions and sum over flavour indeces of partons.} as
\begin{equation}
\label{eq:ct}
d{\sigma}^{CT}=d{\sigma}_{LO}^F\otimes\Sigma^F(q_T/Q)\, d^2{\bf q}_{\, T}\;\;\;\;\;,\;\;\;\;\; \Sigma^F(q_T/Q)
\rightarrow
\sum_{n=1}^\infty
\left(\f{\as}{\pi}\right)^n\sum_{k=1}^{2n}
\Sigma^{F(n;k)} \;\f{Q^2}{q_T^2}\ln^{k-1} \f{Q^2}{q_T^2}  \;\;.
\end{equation}
The function $\Sigma^F(q_T/Q)$ reproduces the singular behaviour of $d{\sigma}^{F+{\rm jets}}$ in the small $\qt$ regime. In this limit it can be expressed  in terms of $\qt$-independent coefficients $\Sigma^{F(n;k)}$. All the single coefficient functions which are taking part in the counter term up to N$^{3}$LO are known analytically, which ensures a correct and precise numerical cancellation of the IR divergencies (from real contributions) associated to the small-$\qt$ limit.
Considering the contribution at $\qt=0$, which restores unitary regarding total cross section, the master formula of the $\qt$-subtraction method is finally:
\begin{equation}
\label{eq:mainqt}
d{\sigma}^{F}_{\rm N^{i}LO}={\cal H}^{F}_{\rm N^{i}LO}\otimes d{\sigma}^{F}_{LO}
+\left[ d{\sigma}^{F+{\rm jets}}_{\rm N^{i-1}LO}-
d{\sigma}^{CT}_{\rm N^{i}LO}\right]\;\; .
\end{equation}
The coefficient ${\cal H}^{F}_{\rm N^{i}LO}$ does not depend on $q_T$ and it is obtained by the N$^{i}$LO truncation of the perturbative function
\begin{equation}
{\cal H}^{F}=1+\f{\as}{\pi}\,
{\cal H}^{F(1)}+\left(\f{\as}{\pi}\right)^2
{\cal H}^{F(2)}+\left(\f{\as}{\pi}\right)^3
{\cal H}^{F(3)}+ \dots \;\;.
\end{equation}
At N$^{3}$LO the hard--virtual functions ${\cal H}^{F(i)}$, with $i = 1,2,3$ are required by the $\qt$-subtraction method. The general structure of ${\cal H}^{F(1)}$ and ${\cal H}^{F(2)}$ (and their specific coefficient functions which are part of them) is explicitly known~\cite{deFlorian:2001zd,Catani:2000vq,Catani:2011kr,Catani:2012qa,Catani:2013tia}. The general structure of ${\cal H}^{F(3)}$ is known \cite{Cieri:2018oms} but some of its ingredients are still missing. Nevertheless, within the $\qt$-subtraction formalism, all these missing coefficients can be inferred for any hard scattering process whose corresponding total cross section is known at N$^{3}$LO. This point is discussed in detail in Sec. 3 of Ref. \cite{Cieri:2018oms}.

\section{The rapidity distribution of the Higgs boson at N$^{\rm3}$LO}
\label{sec:results}
In this section we present our predictions for the Higgs boson rapidity distribution at the LHC, applying the N$^{\rm3}$LO $\qt$-subtraction method presented in Sec.~\ref{sec:forma} for $F = H$. 

We consider Higgs boson production ($M\equiv M_H= 125$~GeV) in proton--proton collisions at a centre-of-mass energy of $\sqrt{s}=13$~TeV in the large-$m_t$ limit ($m_{t}\rightarrow \infty$). In this limit, the production of the Higgs boson is described through an effective gluon-gluon-Higgs boson vertex. Note that we systematically employ the same order in the PDFs (in particular the set \verb|PDF4LHC15_nnlo_mc|~\cite{nnpdf}) for the LO, NLO, NNLO and N$^{3}$LO results presented in this proceeding. The central factorization and renormalization scale is chosen as $\mu \equiv \muR = \muF =  M_H / 2$. The theoretical uncertainty is estimated by varying the default scale choice independently for $\muR$ and $\muF$ by factors of $\{1/2,2\}$ while omitting combinations with $\muR/\muF = 4$ or $1/4$, resulting in the common seven-point variation of scale combinations. 
\begin{figure}[tbh]
\centering
\includegraphics[width=.56\linewidth]{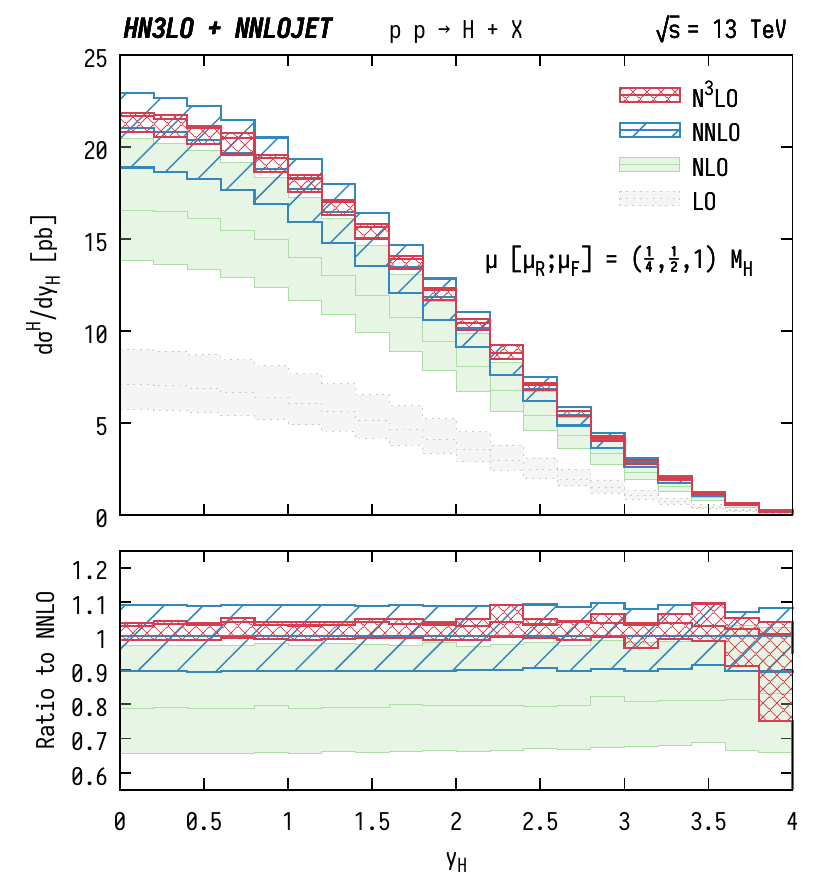}
\caption{\label{fig:yHN3LO}{Rapidity distribution of the Higgs boson computed using the $\qt$-subtraction formalism up to N$^{\rm3}$LO. The seven-point scale variation bands of the LO, NLO, NNLO and N$^{\rm3}$LO results are as follows: LO (pale grey fill), NLO (green fill), NNLO (blue hatched) and N$^{\rm3}$LO (red cross-hatched). The central scale ($\mu=M_{H}/2$) at each perturbative order (except LO) is shown with solid lines. In the lower panel, the ratio to the NNLO prediction is shown. While the bands for the predictions at LO, NLO and NNLO are computed with the seven scales as detailed in the text, the N$^{\rm3}$LO band is obtained after considering also the uncertainties due to the variation of the $\qtcut$ and the  ${\cal H}^{H(3)}$ coefficient in the N$^{\rm3}$LO-only contribution.
}}
\end{figure}
The contributions $d{\sigma}^{H+{\rm jets}}$ in Eq.~\ref{eq:mainqt} are computed with the parton-level event generator \texttt{NNLOJET}, \cite{Chen:2016zka} which provides the necessary infrastructure for the antenna subtraction method up to NNLO.~\cite{Antenna:method} The contributions ${\cal H}^{H}$ and $d{\sigma}^{CT}$ (in Eq.~\ref{eq:mainqt}) are calculated using a new Monte Carlo generator \texttt{HN3LO}.~\cite{leaninprepHN3LO}

Figure~\ref{fig:yHN3LO} shows the rapidity distribution of the Higgs boson at LO (pale grey fill), NLO (green fill), NNLO (blue hatched) and N$^{\rm3}$LO (red cross-hatched).  The central scale ($\mu=M_{H}/2$) is shown as a solid line while the bands correspond to the envelope of seven-point scale variation. At N$^{\rm3}$LO, the band additionally includes the uncertainties due to $\qtcut$ and ${\cal H}^{H(3)}$ as described in Sec.~4.2 of Ref. \cite{Cieri:2018oms} Going from LO to NNLO, the scale $\mu=M_{H}/2$ is always at the center of the respective scale variation band in Fig.~\ref{fig:yHN3LO}.  The central prediction at N$^{\rm3}$LO, on the other hand, almost coincides with the upper edge of the band, as was already observed for the total cross section~\cite{Anastasiou:2015ema,Mistlberger:2018etf}, see Table~2 and Fig.~3 of Ref. \cite{Cieri:2018oms} Figure \ref{fig:yHN3LO} shows a substantial reduction in the size of the scale variation band at N$^{\rm3}$LO, both in the total cross section and in differential distributions. Comparing Fig.~\ref{fig:yHN3LO} with the results obtained in Ref.~\cite{Dulat:2018bfe} we observe very good agreement between the two calculations.

\section{Conclusions and outlook}
\label{Sec:conclu}

In this proceeding we have presented the extension of the $\qt$-subtraction method at N$^{\rm3}$LO applied (for first time) to the rapidity distribution of the Higgs boson at the LHC. We calculate the $y_H$ distribution at N$^{\rm3}$LO employing a seven-point scale variation and carefully assess systematic errors arising form different $\qtcut$ and the approximation made on ${\cal H}^{H(3)}$. Compared to the NNLO $y_H$ distributions, we observe a large reduction of theory uncertainties by more than $50\%$ at N$^{\rm3}$LO. The scale variation band at N$^{\rm3}$LO stays within the NNLO band with a flat $K$-factor of about $1.034$ in the central rapidity region ($|y_H|\leq3.6$). Both the systematic error analysis and the phenomenological predictions confirm that our calculations at N$^{\rm3}$LO using $\qt$-subtraction formalism are well under control. The approximation related to some of the coefficients functions of ${\cal H}^{H(3)}$  in our approach, can be easily replaced by the full analytical results once available.

\section*{References}

\end{document}